\documentclass[letterpaper, 10 pt, conference]{ieeeconf}  

\IEEEoverridecommandlockouts                              

\usepackage{graphicx}
\usepackage{float}
\usepackage[caption=false]{subfig}
\usepackage{mathptmx} 
\usepackage{times} 
\usepackage{booktabs}
\usepackage{tabularx}
\usepackage{multirow}
\usepackage{marvosym}
\usepackage{stfloats}
\usepackage{array}
\usepackage{soul, color, xcolor}
\usepackage{hyperref}
\usepackage{fancyhdr}
\title{\LARGE \bf
RISEE: A Highly Interactive Naturalistic Driving Trajectories Dataset with Human Subjective Risk Perception and Eye-tracking Information*
}

\author{Xinzheng Wu, Junyi Chen\textsuperscript{\Letter}, Peiyi Wang, Shunxiang Chen, Haolan Meng, and Yong Shen
\thanks{*This work was supported by National Key R\&D Program of China (2021YFB2501205), National Natural Science Foundation of China (52232015) and Fundamental Research Funds for the Central Universities.}
\thanks{All authors are with the School of Automotive Studies, Tongji University, Shanghai 201804, China. Corresponding E-mail:
        {\tt\small chenjunyi@tongji.edu.cn}}%
\thanks{$^{1}${\tt\small \href{https://ivtest-lab.github.io/RISEE_dataset/}{https://ivtest-lab.github.io/RISEE\_dataset/}} }%
}

\begin{document}

\maketitle
\thispagestyle{fancy}
\pagestyle{empty}

\begin{abstract}

In the research and development (R\&D) and verification and validation (V\&V) phases of autonomous driving decision-making and planning systems, it is necessary to integrate human factors to achieve decision-making and evaluation that align with human cognition. However, most existing datasets primarily focus on vehicle motion states and trajectories, neglecting human-related information. In addition, current naturalistic driving datasets lack sufficient safety-critical scenarios while simulated datasets suffer from low authenticity. To address these issues, this paper constructs the Risk-Informed Subjective Evaluation and Eye-tracking (RISEE) dataset which specifically contains human subjective evaluations and eye-tracking data apart from regular naturalistic driving trajectories. By leveraging the complementary advantages of drone-based (high realism and extensive scenario coverage) and simulation-based (high safety and reproducibility) data collection methods, we first conduct drone-based traffic video recording at a highway ramp merging area. After that, the manually selected highly interactive scenarios are reconstructed in simulation software, and drivers' first-person view (FPV) videos are generated, which are then viewed and evaluated by recruited participants. During the video viewing process, participants' eye-tracking data is collected. After data processing and filtering, 3567 valid subjective risk ratings from 101 participants across 179 scenarios are retained, along with 2045 qualified eye-tracking data segments. The collected data and examples of the generated FPV videos are available in our website$^{1}$.

\end{abstract}

\section{INTRODUCTION}

Recent years have witnessed the rapid development of autonomous driving technology. As the "brain" of autonomous vehicles, the decision-making and planning (D\&P) system plays a crucial role. In the research and development (R\&D) phase of D\&P systems, with recent advancements in deep learning technologies, learning-based algorithms have been extensively studied and proven capable of handling complex driving tasks in diverse environments \cite{zhang2025decisionmaking}. However, training a well-informed algorithm requires a large amount of high-quality data.

To address this issue, vast amounts of vehicle-side data have been collected by both manufacturers and research institutions, such as the Waymo dataset \cite{sun2020scalability} and the nuScenes dataset \cite{caesar2020nuscenes}. However, since the number of multi-sensor-equipped collection vehicles is limited, achieving diverse scenario coverage is prohibitively expensive, not to mention including collision scenarios due to safety and ethical constraints. Therefore, from another perspective, many researchers use camera-equipped drones to capture the traffic from a bird's eye view and extract datasets of vehicle trajectories, such as the HighD \cite{krajewski2018highd} and SIND \cite{xu2022sind} datasets. Nevertheless, the above-mentioned datasets only focus on vehicle motion states but lack human factors such as subjective risk perception and physiological data, which limits the algorithms' ability to make decisions aligned with human cognition.

Further, in the verification and validation (V\&V) phase of D\&P systems, driving datasets with human factors are also essential to extract and generate critical testing scenarios \cite{wu2025make}, construct human baseline driver model \cite{mattas2022drivera} and evaluate safety and intelligence performance \cite{you2025comprehensive}, etc.

Aiming at constructing human-factor-integrated driving datasets, numerous studies have conducted real-world or simulation experiments to capture drivers/passengers' subjective risk perception and physiological data. For instance, Ke et al. \cite{ke2024d2e} collected human drivers' driving behavior and physiological data using a 3-DoF (three-degree-of-freedom) driving simulator across 12 custom-designed scenarios, while additionally recruiting 40 volunteers as expert evaluators to obtain human subjective evaluations for each scenario. Meng et al. \cite{meng2024study} collected passengers' physiological signals (e.g., eye-tracking data, electrodermal activity) through real-world experiments, while using a slider device to capture their real-time subjective feedback (e.g., perceived comfort or risk levels). You et al. \cite{you2025comprehensive} first conducted real-world experiments and then obtained subjective evaluations from both drivers and passengers by asking participants to review the recorded videos during experiments in the post-experiment interviews.

However, both simulation experiments and real-world experiments still exhibit inherent limitations. For simulation experiments, the realism of scenarios (including both the authenticity of surrounding vehicles' behaviors and the environmental fidelity), alongside the realism of driving experience (e.g., force feedback, audiovisual cues, control latency) remain critical concerns. In contrast, real-world experiments exhibit high fidelity but they are constrained by safety requirements, limiting the interactivity and criticality of the collected scenarios, which results in a scarcity of safety-critical scenarios.

Towards addressing the aforementioned issues, this paper constructs a \textbf{R}isk-\textbf{I}nformed \textbf{S}ubjective \textbf{E}valuation and \textbf{E}ye-tracking (RISEE) dataset that includes both naturalistic driving trajectories and human factors. Taking advantage of the low cost and high scenario coverage of drone-based data collection methods, we first conduct video recordings at a highway merging zone. Subsequently, naturalistic driving scenarios are extracted and highly interactive scenarios are then manually selected. After that, high-fidelity driver's first-person view (FPV) videos are reconstructed by simulation with optimizations to road surface textures, traffic infrastructure (e.g., lane markings, signage), environmental elements (e.g., static objects, weather effects), and acoustic feedback (e.g., engine noise, tire friction). Finally, volunteers are recruited to view these videos in a driving simulator, providing subjective risk perception scores. Throughout this process, their eye-tracking data (e.g., gaze fixation points, saccadic movements) is synchronously recorded using a head-mounted eye tracker.

Compared with existing human-factor-integrated driving datasets, the contributions of RISEE are as follows:

\begin{itemize}
    \item Highly interactive naturalistic driving scenarios: The RISSE dataset contains 179 highly interactive naturalistic driving scenarios, encompassing multiple interaction patterns such as car-following, cut-in, overtaking, and ramp merging, along with diverse driving risk levels ranging from safe to near-crash events. Since all the scenarios are derived from real-world occurrences, their plausibility and authenticity are inherently guaranteed.
    \item High-fidelity FPV videos: High-fidelity driver's FPV videos are reconstructed based on the recorded trajectories. To ensure the realism of the driving experience, both the external vehicle environment (including road surface textures, traffic infrastructure, environmental elements, and acoustic feedback) and the vehicle interior (including the dashboard, turn signals, rearview mirrors) are carefully designed and rendered.
    \item Both subjective and objective risk information with eye-tracking data: A total of 102 volunteers are recruited, and their subjective risk ratings for the scenarios along with eye-tracking data are collected. Additionally, based on the vehicle kinematic states, the driving risk for each scenario is assessed using specific risk indicators and provided as scenarios' objective risk in the dataset.
\end{itemize}

\begin{figure*}[b]
    \centering
    \includegraphics[width=.95\textwidth]{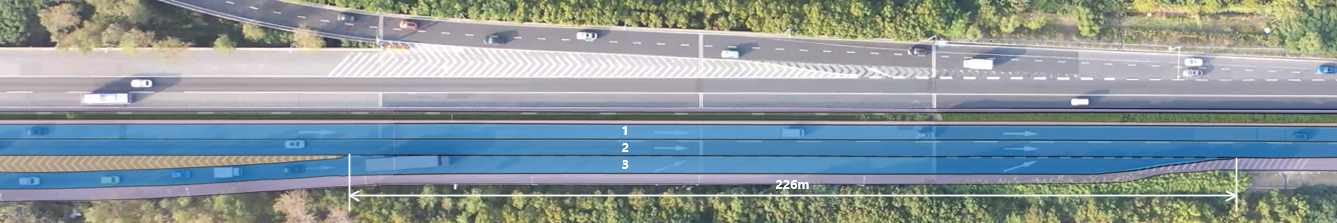}
    \caption{The road network of the recording site.}
    \label{recodingSite}
\end{figure*}

The remainder of the paper is organized as follows: Section \ref{sec2} introduces the data acquisition method, including naturalistic driving trajectory extraction, drivers’ PFV videos generation, and human factor data collection. In Section \ref{sec3}, the collected data is processed and analyzed, with discussions from both driver characteristics and scenario characteristics perspectives. Section \ref{sec4} illustrates potential applications of the proposed dataset. Section \ref{sec5} concludes the paper.

\section{DATA ACQUISITION METHOD} \label{sec2}

\subsection{Drone-based Traffic Data Recording and Processing}

To cover as many interaction patterns as possible, we select a highway on-ramp merging section as the recording site, where there exist both merging behaviors from the ramp and various main-road behaviors including car-following, overtaking, and cut-in maneuvers. Specifically, the road section in RISEE is the entrance of Jiasong Middle Road of the G50 Shanghai-Chongqing Expressway in China, including two lanes in the main road and an on-ramp with a 226-meter acceleration lane, as shown in Fig. \ref{recodingSite}.

To record the traffic data, a DJI Mavic 2 Pro drone is deployed, maintaining a stable hover at an altitude of 300 meters and conducting a 4-hour continuous traffic data collection. After recording, the lane information is first identified using ArcMap \cite{arcmap} and converted into OpenDrive \cite{asam} format for further simulation reconstruction. Then, a convolutional neural network (CNN)-based method is employed for vehicle detection and bounding box construction. Since the recognition algorithm is not the focus of this study, we utilize the established YOLOv5 \cite{yolov5} architecture as our detection framework. Next, highly interactive scenario segments are manually selected for subsequent simulation reconstruction. It should be noted that, to mitigate the computational and rendering load during reconstruction, only interaction-relevant vehicles are retained in the scenarios. Finally, a total of 179 highly interactive scenarios are extracted, with vehicle trajectory information in the scenarios (including position, orientation, speed, acceleration, etc.) stored in CSV files. Moreover, in each scenario file, the first vehicle is designated as the ego vehicle, serving as the reference for driver's perspective in the subsequent FPV videos generation process. The number of vehicles in the 179 extracted scenarios ranges from 2 to 7, with the detailed distribution presented Table. \ref{tab1}.

\begin{table}[ht]
\renewcommand{\arraystretch}{1.1}
\caption{Vehicle Number Distribution in the Extracted Scenarios.\label{tab1}}
\centering
\begin{tabular}{ m{0.3\linewidth}<{\centering}  m{0.3\linewidth}<{\centering} }
\toprule
\textbf{Number of Vehicles} & \textbf{Number of Scenarios} \\
\midrule
2 & 12   \\
3 & 32   \\
4 & 73   \\
5 & 37   \\
6 & 12   \\
7 & 13   \\ \hline
\textbf{Total} & \textbf{179} \\
\bottomrule
\end{tabular}
\end{table}

\subsection{Simulation Reconstruction and Driver's FPV Videos Generation}

\begin{figure*}[b]
    \centering
    \includegraphics[width=.95\textwidth]{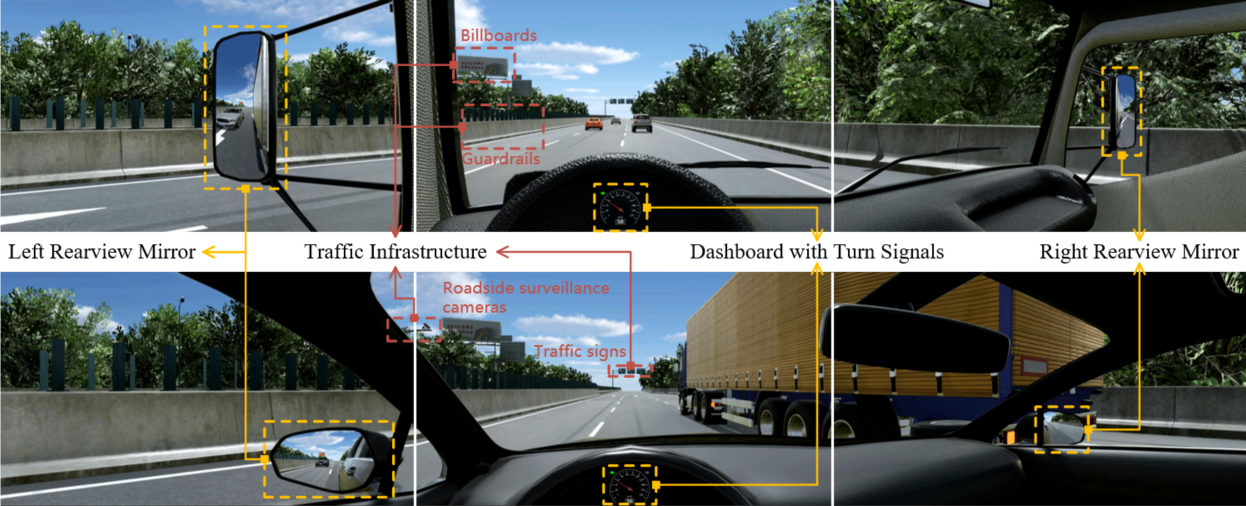}
    \caption{Illustrations of the driver’s FPV perspectives. The upper presents the truck's viewpoint, while the lower demonstrates the sedan's viewpoint.}
    \label{FPV_video}
\end{figure*}

In this paper, the simulation software SimOne \cite{simone} is chosen to reconstruct the scenarios. This software uses a graphics rendering method that integrates 3D Gaussian splatting, which can ensure a basic level of realism. Generally, by importing vehicle trajectory files into the simulation software, the simulation reconstruction of scenarios can be achieved. Furthermore, in addition to the built-in scenario objects provided by the simulation software, we have specially implemented additional visual optimizations to ensure an authentic driving experience. Fig. \ref{FPV_video} illustrates the simulation-reconstructed driver's FPV perspective.

As can be seen in the figure, according to the different types of ego vehicles, FPV perspectives for sedans and trucks are separately generated. More in detail, each FPV perspective includes the driver's forward view, left view, and right view. Based on the normal human field of view (FOV), we set the wide-angle of cameras in every direction to 60$^\circ$ after evaluating exterior object distortion under varying wide-angle camera configurations. By adjusting the position coordinates of the cameras to allow their images to be stitched together without overlapping, FPV videos with a horizontal FOV of 180$^\circ$ are generated. 

When it comes to the visual optimizations, for the vehicle's external environment, we have added abundant traffic infrastructure to match the actual situation of Chinese highways, including guardrails, billboards, traffic signs and roadside surveillance cameras. For the vehicle interior, we have additionally rendered the dashboard and the perspectives of the rearview mirrors on both sides, thereby enabling human evaluators to have a better understanding of the vehicle current velocity and driving environment. At the same time, turn signals are also displayed in the dashboard, thereby demonstrating the vehicle's lane-changing intention. Since the entire trajectory of the vehicle is known, the lane-changing intention of the vehicle at the current moment can be determined by the actual motion state at the next moment. Specially, if the vehicle is located on the ramp or acceleration lane (Lane No.3 in Fig.\ref{recodingSite}), then we stipulate that the left turn signal remains activated, because the vehicle ultimately needs to complete ramp merging.

In addition, to obtain a more immersive driving experience, acoustic feedback including engine noise and tire noise of all vehicles is generated. Specifically, according to the relative positions and relative speeds between vehicles, stereo sound is generated based on SumoSound \cite{patmalcolm91}. Examples of the driver's FPV videos can be found on our website.

\subsection{Human Subjective Evaluation and Eye-tracking Data Collection} \label{sec2-c}

In this paper, we recruit volunteers as human evaluators through a questionnaire. The questionnaire collects volunteers' basic personal information and driving-related information, while assessing their experimental compatibilities. The experimental compatibility assessment includes questions regarding whether participants wear glasses, have heart disease, might experience nervousness when wearing experimental equipment, or have previously participated in similar experiments. Through the questionnaire screening process, a total of 102 volunteers are recruited for data collection. The detailed demographic and driving-related information of the participants is shown in Table \ref{tab2}.

From Table \ref{tab2}, we can find that our participants come from diverse age groups and exhibit varying characteristics in terms of driving experience, driving frequency, and driving ability. It is worth mentioning that 8 participants do not have a driver’s license (their driving-related characteristics are labeled as None). We believe feedback from individuals without a driver’s license is still valuable, since future high-level autonomous driving systems may not necessarily require human drivers to hold a license.

\begin{table}[t]
\renewcommand{\arraystretch}{1.1}
\caption{Information of the Participants.\label{tab2}}
\centering
\begin{tabular}{ m{0.15\linewidth}  m{0.33\linewidth} m{0.15\linewidth} m{0.15\linewidth}}
\toprule
\textbf{Information} & \textbf{Characteristics} & \textbf{Number} & \textbf{Ratio}\\
\midrule
\multirow{2}{*}{Gender} & Male  &  78 & 76.5\% \\
~ & Female & 24 & 23.5\% \\ \hline
\multirow{4}{*}{Age} & 19-25  &  79 & 77.5\% \\
~ & 26-30 & 15 & 14.7\% \\
~ & 31-45 & 4 & 3.9\% \\
~ & 46-55 & 4 & 3.9\% \\ \hline
\multirow{4}{.8\linewidth}{Driving Years} & None  &  8 & 7.8\% \\
~ & 0-5 & 73 & 71.6\% \\
~ & 6-10 & 16 & 15.7\% \\
~ & 10+ & 5 & 4.9\% \\ \hline
\multirow{5}{.8\linewidth}{Driving Frequency} & None  &  8 & 7.8\% \\
~ & Less than once a month & 47 & 46.1\% \\
~ & At least once a month & 22 & 21.6\% \\
~ & At least once a week & 14 & 13.7\% \\ 
~ & At least once a day & 11 & 10.8\% \\ \hline
\multirow{5}{\linewidth}{Self-assessment of Driving Ability} & None  &  8 & 7.8\% \\
~ & Novice & 19 & 18.6\% \\
~ & Intermediate & 26 & 25.5\% \\
~ & Proficient & 27 & 26.5\% \\ 
~ & Expert & 22 & 21.6\% \\
\bottomrule
\end{tabular}
\end{table}

\begin{figure*}[b]
    \centering
    \includegraphics[width=.95\textwidth]{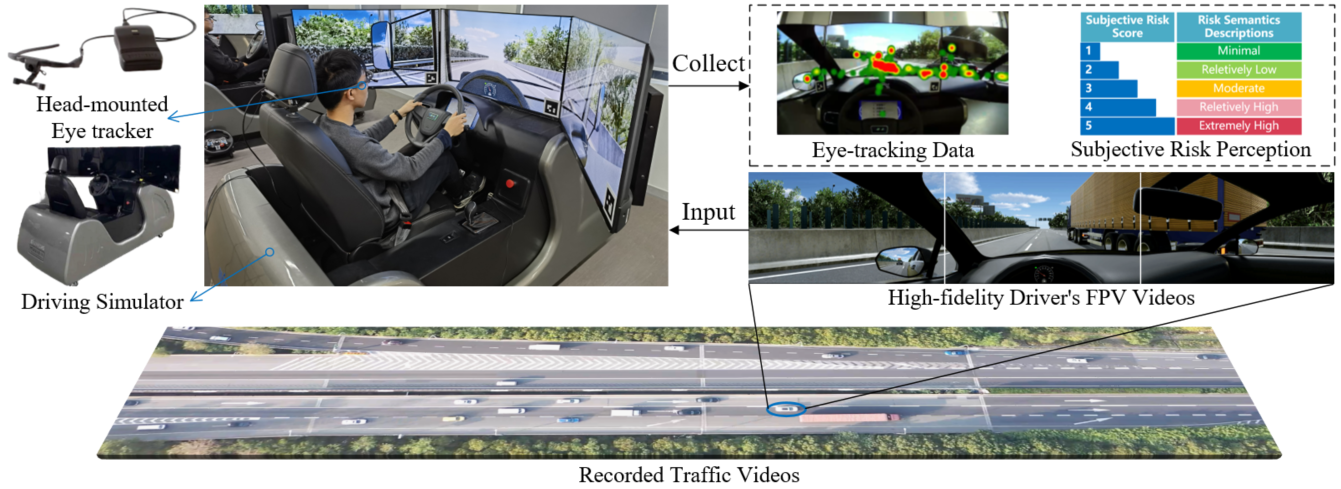}
    \caption{Illustration of human data collection pipline.}
    \label{data_collection_pipline}
\end{figure*}

Before the data collection, each participant is fully informed of potential risks and discomforts, privacy protections, and their right to withdraw freely from the study. In addition, participants are required to complete a questionnaire to assess their sensitivity of risk perception. This questionnaire is adapted from \cite{ma2010safetya}, which primarily captures participants' worries about traffic risks, their perceived likelihood of accidents, and their concerns about traffic risks and being victimized, with each item measured on five-point Likert scales ranging from “strongly disagree” to “strongly agree”. The pipeline of the human data collection is illustrated in Fig. \ref{data_collection_pipline}. 

As shown in Fig. \ref{data_collection_pipline}, participants are asked to sit in a driving simulator to view the generated driver’s FPV videos, with a head-mounted eye tracker capturing their eye-tracking data. In this study, although no manual control of the vehicle is required, the driving simulator is still used to preserve real driving experience during video viewing. The three screens of the driving simulator are set at a 120-degree angle, matching the FOV configuration used during video generation. At the same time, stereo audio is delivered through headrest-mounted speakers to enhance immersion. Notably, in preliminary experiments we find that physiological signals such as ECG and EDA can't respond quickly to scenario changes due to the short duration of each scenario (15-20 seconds). Therefore, only eye-tracking data is collected.

Prior to the formal video viewing sessions, each participant will first watch two baseline videos (with sedan's and truck's perspective, respectively) which contain no other vehicles to acclimate to the viewing environment. Subsequently, one typical safe scenario and one typical critical scenario are selected and presented, thereby calibrating participants’ risk expectations (i.e., all the viewed scenarios’ risk levels are bracketed between these two reference extremes).
Finally, the remaining 177 scenarios are divided into 10 groups, each containing 17 to 18 scenarios. To mitigate possible evaluation mistakes caused by fatigue, each participant is assigned to watch only two groups of videos, with each video played twice to ensure full comprehension of the scenario. Moreover, the selection of video groups is uniform to ensure that each video is viewed in similar frequency. By doing so, each video could be viewed by at least 20 different participants. After viewing each video, participants will report their subjective risk perception, measured via a 5-point Likert scale ranging from “minimal risk” to “extremely high risk”, as shown in Fig. \ref{data_collection_pipline}. Note that to ensure unbiased risk-level distribution across groups (as the dataset includes both safe and critical scenarios), we first calculate the DNDA metric for each scenario as an objective risk value. DNDA is a normalized risk indicator based on drivable area \cite{wu2022risk}. The closer its value is to 1, the more critical the scenario is.
Based on these values, safe and critical scenarios are evenly allocated to each group.

\section{DATA ANALYSIS} \label{sec3}

\subsection{Data Processing and Screening}
Since eye-tracking data is continuously recorded during the data collection process, containing redundant information during scenario video transitions, the eye-tracking data is first segmented and aligned with the scenario data. At the same time, to ensure data quality, internal consistency checks are performed on participants' subjective risk perception feedback to identify and exclude careless or insincere ratings. More in detail, within each video group, duplicate scenarios are inserted at distant intervals. Participants exhibiting inconsistent subjective risk perceptions (defined as a rating discrepancy exceeding 1 point) between repeated scenarios are identified, and all their feedback within that video group are removed from subsequent analyses. Further, due to issues such as device disconnections, intermittent latency, and excessive timestamp inaccuracies in the eye-tracking equipment during the experiment, not all collected eye-tracking data is valid. Ultimately, a total of 3567 valid subjective risk perception ratings of 101 participants are retained (One participant doesn't pass the consistency checks in both scenario groups), accompanied by 2045 valid eye-tracking data segments.

\subsection{Overall Distribution of Scenario Subjective and Objective Risks} \label{sec3-b}
\begin{figure}[t]%
    \centering
    \subfloat[Scenario distribution of TTC calculation results]{
        \label{TTC_dis}
        \includegraphics[width=.8\linewidth]{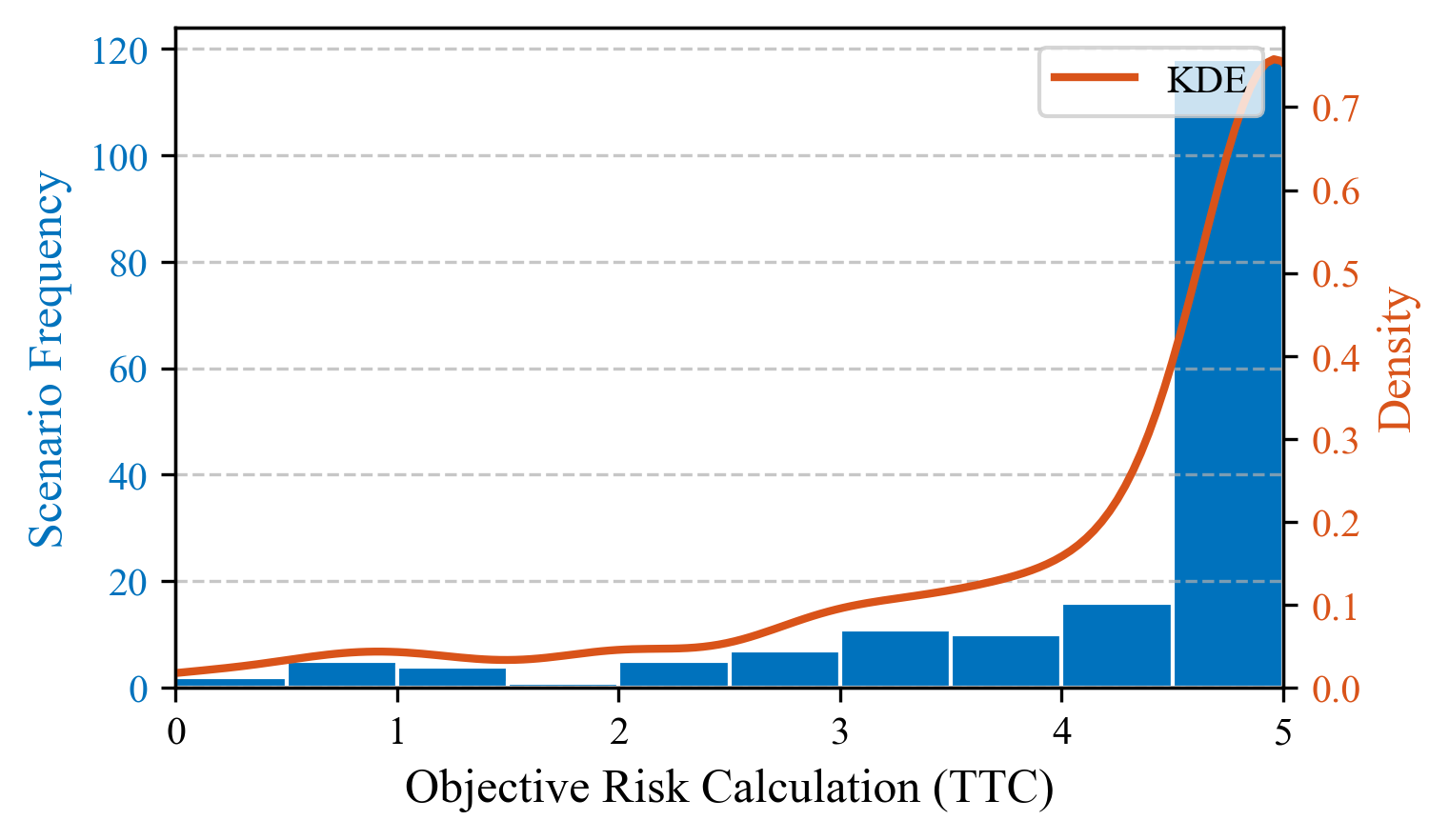}
        } \vspace{0.01cm}
    \subfloat[Scenario distribution of DNDA calculation results]{
        \label{DNDA_dis}
        \includegraphics[width=.8\linewidth]{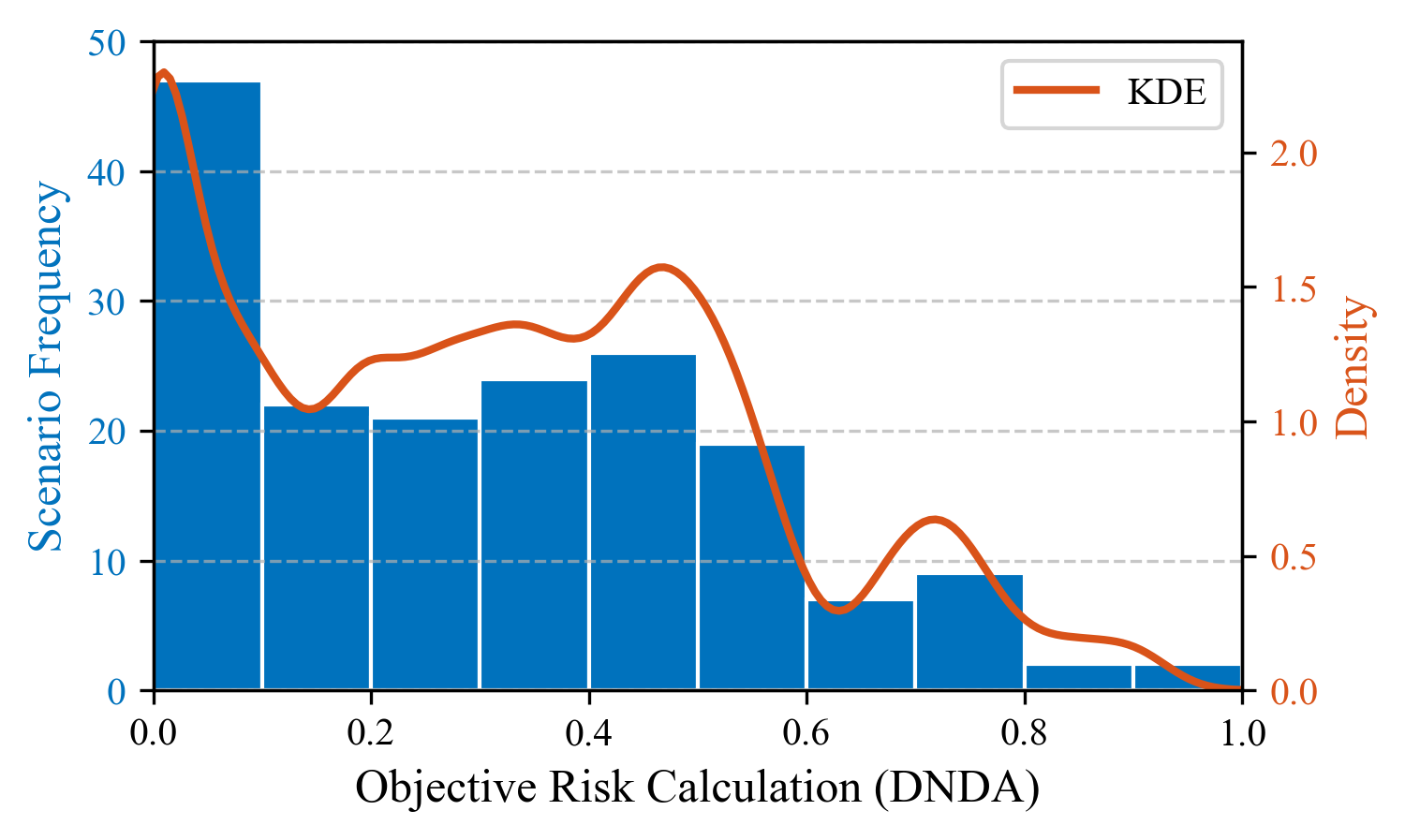}
        } \vspace{0.01cm}
    \subfloat[Scenario distribution of subjective risk perception values]{
        \label{SUB_dis}
        \includegraphics[width=.8\linewidth]{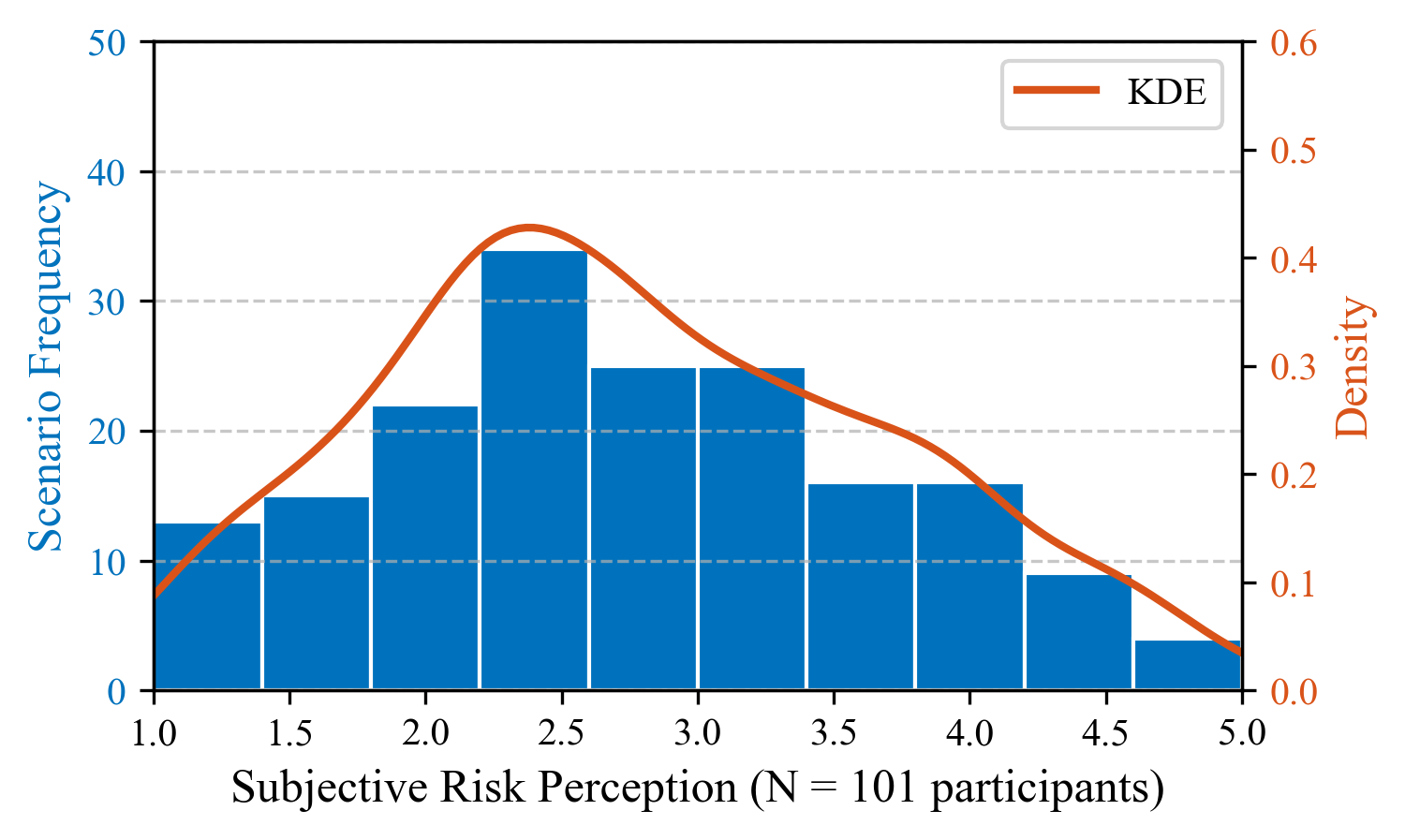}
        }
    \caption{Illustrations of the overall subjective and objective risk distributions.
    }
    \label{scenarioDis}
\end{figure}

As previously described, each scenario is viewed and rated by multiple participants for subjective risk perception. Therefore, the mean value across participants is calculated as the subjective risk value for each scenario. Concurrently, the DNDA metric and time to collision (TTC) metric for each scenario is computed, with the maximum DNDA value and the minimum TTC value observed during the scenario serving as the objective risk value. It is worth noting that when calculating the TTC, in addition to computing the TTC between the ego vehicle and the preceding vehicle, the TTC of the following vehicle behind ego vehicle is also calculated to assess the risk of being rear-ended. Finally, the frequency distribution histograms of subjective and objective risks across the 179 scenarios are presented in Fig. \ref{scenarioDis}. Moreover, to better visualize the distribution of subjective and objective risks across various scenarios, Kernel Density Estimation (KDE) is applied to generate probability density curves. The DNDA and TTC risk values for all scenarios are also provided in the RISEE dataset.

As can be seen from the figure, there are significant differences in the distributions of subjective and objective risks. Since TTC values exceeding 5 seconds are generally considered to represent relatively safe scenarios, scenarios with TTC $>$ 5s are all categorized as TTC $=$ 5s for statistical calculation. According to the TTC and DNDA calculation results, the objective risks of the 179 scenarios are mostly at low to moderate levels, while based on human subjective perception, the subjective risks of scenarios are concentrated at moderate levels. We believe this is because the objective risk indicators are entirely based on the vehicle's kinematic states for risk assessment. In contrast, human subjective risk perception incorporates more factors beyond the scope of objective risk indicators (such as the vehicle approaching guardrails or crossing lane markings, and intentions of other vehicles to change lanes based on turn signal information), thus resulting in higher risk perception.

\begin{figure}[t]
    \centering
    \includegraphics[width=.6\linewidth]{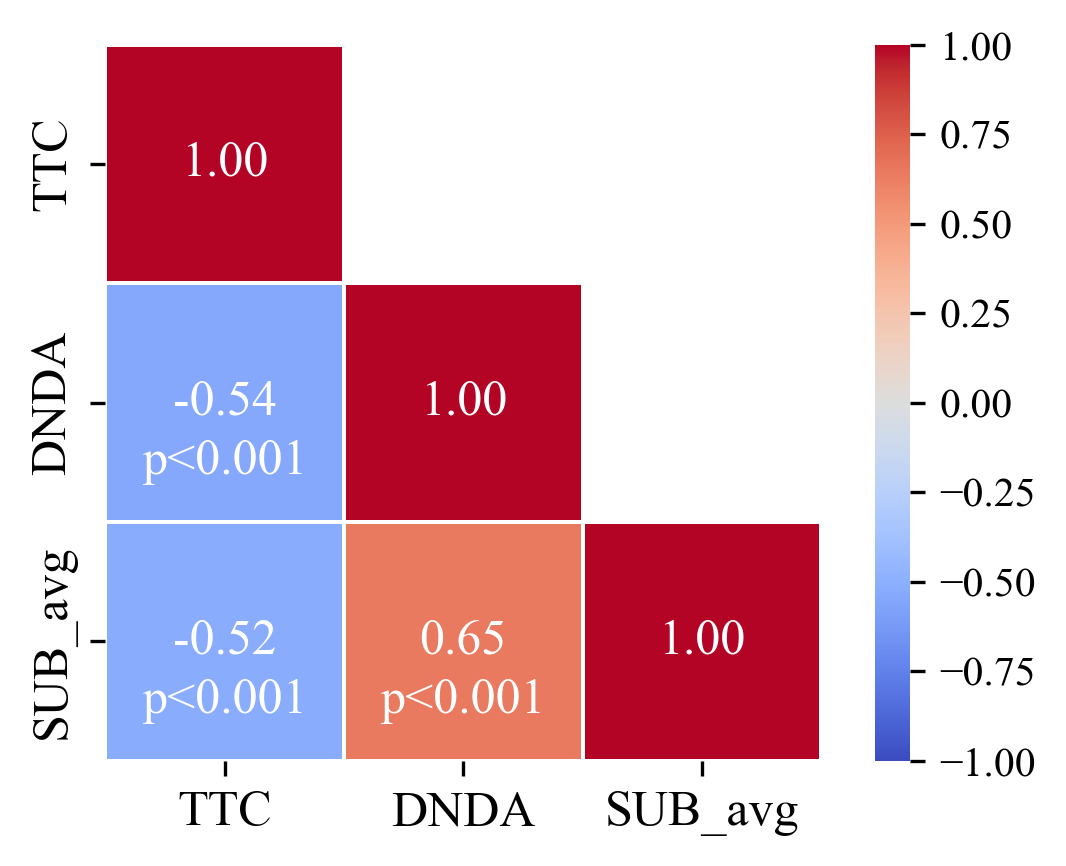}
    \caption{Heatmap of the Spearman correlation coefficients among the three risk distributions.}
    \label{correlation_heatmap}
\end{figure}

Furthermore, we have calculated the Spearman correlation coefficients among the three risk distributions. As shown in Fig. \ref{correlation_heatmap}, while demonstrating high statistical significance (p$<$0.001), the results only exhibit a certain degree of correlations between these risk distributions, indicating the differences between subjective and objective risks. This analysis once again demonstrates the irreplaceability of subjective risks and the necessity of collecting human factors.

\subsection{Effects of the Drivers' Sensitivity of Risk Perception} \label{sec3-c}
\begin{figure}[b]
    \centering
    \includegraphics[width=\linewidth]{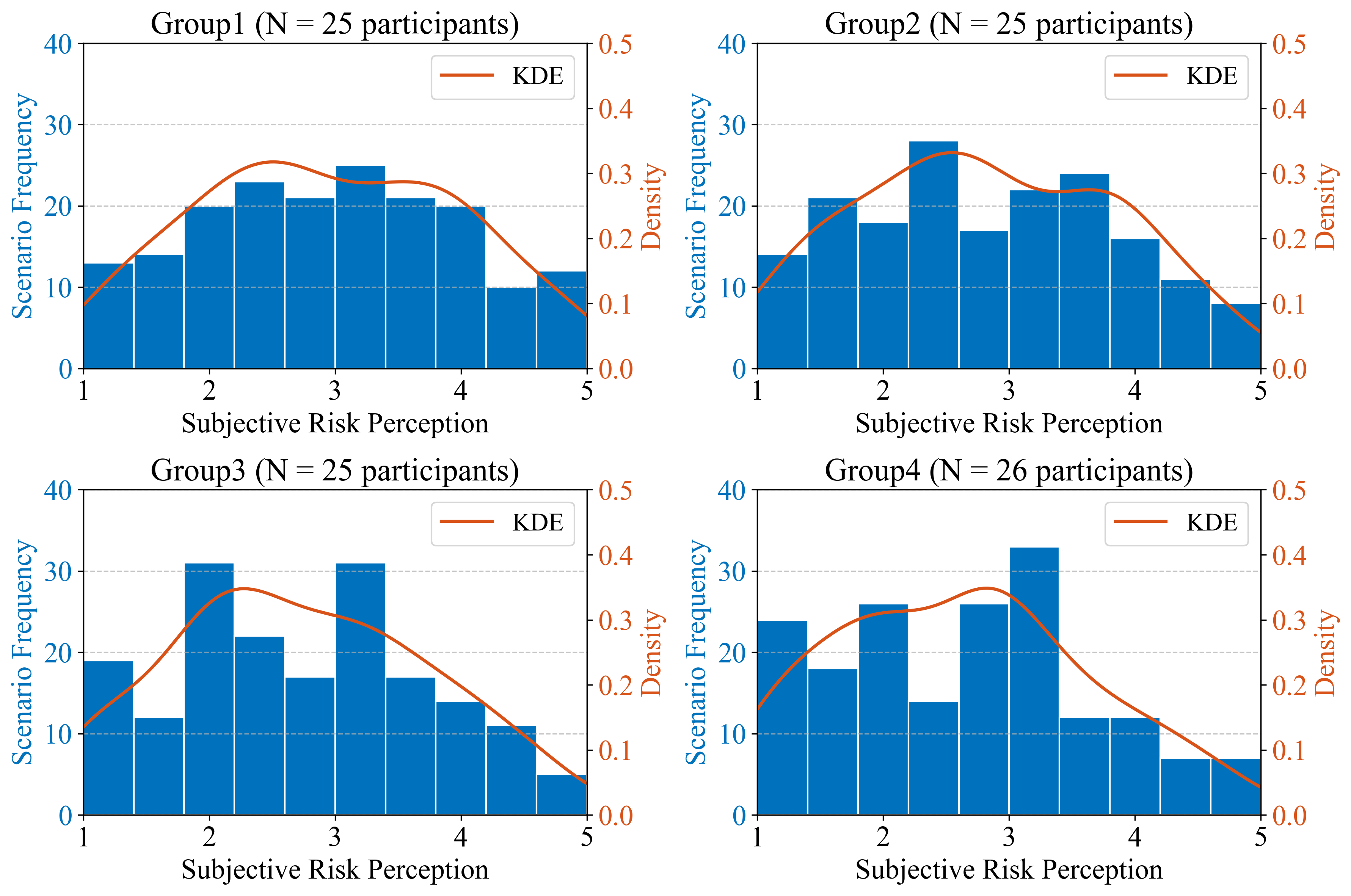}
    \caption{Distribution of scenario subjective risk perception results among groups of different risk sensitivity level. From Group1 to Group4, participants' risk sensitivity level becomes progressively lower, indicating that their tolerance for risk increases correspondingly.}
    \label{risk_sensitivity}
\end{figure}

As mentioned in Section \ref{sec2-c}, we have additionally used a questionnaire to evaluate participants' sensitivity of risk perception. Based on these results, we divide all participants into four groups and study their rating patterns. Specifically, according to the questionnaire results, we first calculate each participant's risk sensitivity score. Since the direction of risk correlation varies across questions (i.e., "strongly agree" indicated high risk sensitivity in some questions but low risk sensitivity in others), we adjust the results to ensure that scores positively correlated with risk sensitivity. After that, participants are divided into four groups of roughly equal size based on their adjusted risk sensitivity scores, ranked from highest to lowest. Thanks to the uniform and unbiased selection of video groups during data collection, all four groups with different risk sensitivity levels cover all the 179 scenarios, and their subjective risk perception distributions across these scenarios are shown in Fig. \ref{risk_sensitivity}.

As demonstrated in Fig. \ref{risk_sensitivity}, with participants' risk perception sensitivity decreasing across Groups 1-4, there is a corresponding reduction in scenarios rated as high-risk. Notably, compared to Group 1 and Group 2, participants in Group 3 and Group 4 have higher risk tolerance, so their scores are concentrated in the 1-3 point range (with lower scores indicating lower risk perception). At the same time, despite differences in risk perception sensitivity, certain scenarios are consistently perceived as high-risk across all groups, indicating that participants reach consensus on extreme scenarios. The above analysis shows that this dataset contains rich driver samples, while also demonstrating potential future applications for personalized decision-making training targeting drivers with different risk perception sensitivity levels.

\begin{figure}[t]
    \centering
    \includegraphics[width=.8\linewidth]{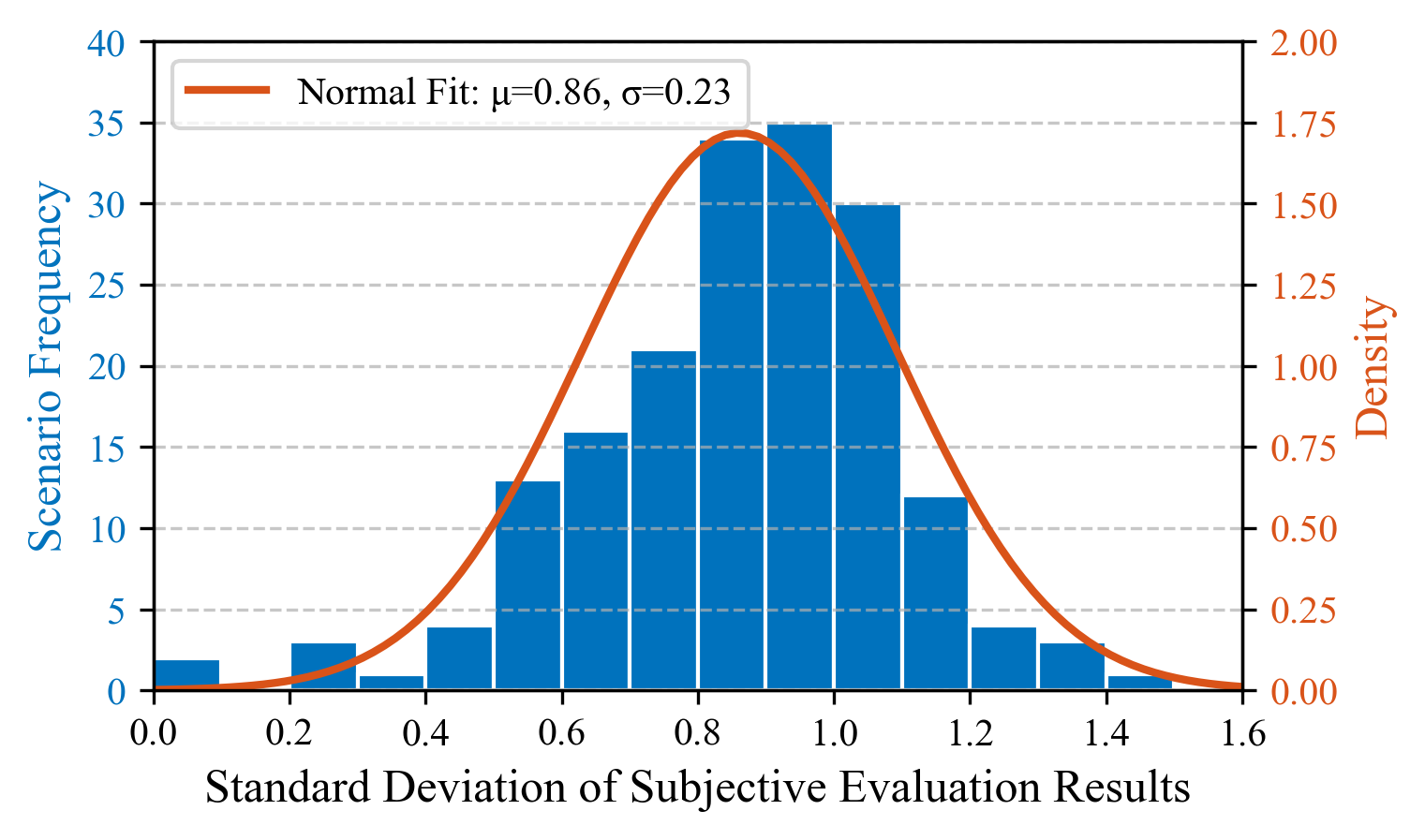}
    \caption{Distribution of the standard deviation of subjective evaluation results.}
    \label{SD_dis}
\end{figure}

\subsection{Analysis on the Subjective Evaluation Consistency of Scenarios}

In order to analyze the consistency and differences in subjective evaluations of the same scenarios by different participants, the standard deviation (SD) of subjective evaluation results for all 179 scenarios is calculated and their distribution is demonstrated in Fig. \ref{SD_dis}, with the orange curve fitted to a normal distribution. As can be seen from Fig. \ref{SD_dis}, for most scenarios, the standard deviations of participants' subjective evaluation results range between 0.6 and 1.2, falling within the normal range of subjective risk perception variations among participants. However, a small number of scenarios exhibit standard deviations of 0 or exceeded 1.3, indicating that participants either reach complete consensus or express divergent perceptions in these specific scenarios.

\begin{figure}[b]
    \centering
    \includegraphics[width=.9\linewidth]{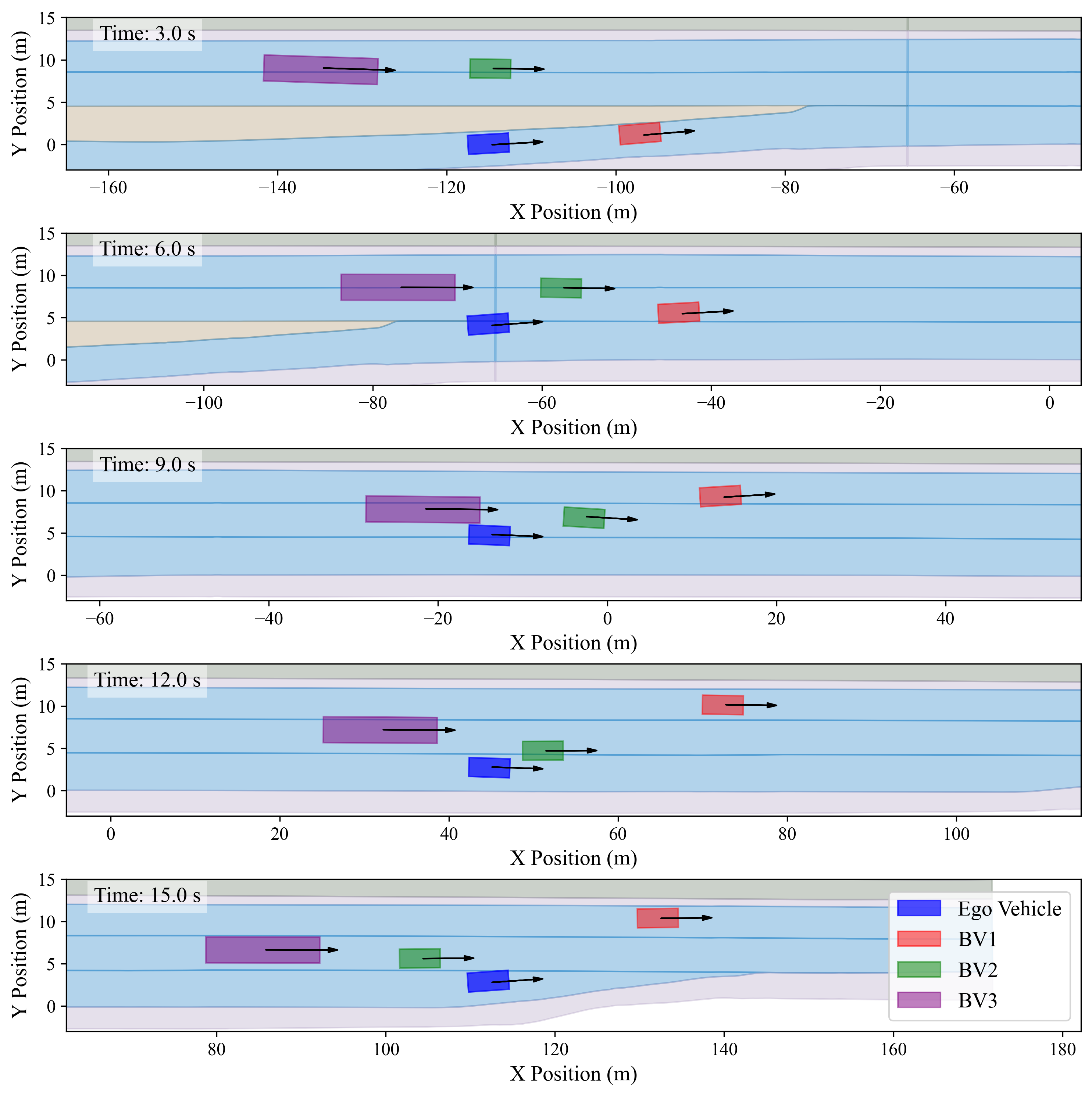}
    \caption{Visualization of the scenario evaluated as “extremely high risk” by all participants. (SD=0, minTTC=0.57s, maxDNDA=0.68)}
    \label{critical_sce}
\end{figure}

To further investigate the underlying reasons behind these phenomena of high consensus and high divergence, we conduct case studies on the corresponding scenarios. By examining the two scenarios with a standard deviation of 0, we observe that they either exhibit obviously safe situations (car-following scenarios where the ego vehicle acts as the leading car maintaining a constant speed) or distinctly dangerous conditions, which is visualized in Fig. \ref{critical_sce} to facilitate better comprehension. As shown in Fig. \ref{critical_sce}, the ego vehicle, which is the reference of the driver's FPV perspective, initially merges from the ramp onto the main road. Due to its aggressive lane-changing maneuver, it nearly collides with BV3 traveling normally on the main road at 9.0s. After that, the ego vehicle accelerates rightward, attempting to overtake BV2 via the acceleration lane, during which process it again nearly collides with BV2 (as depicted at the 12.0s in the figure). Ultimately, the ego vehicle narrowly completes the merging maneuver at the end of the acceleration lane. This scenario is so critical that all participants, no matter their risk perception sensitivity levels, evaluate it as "extremely high risk".

\begin{figure}[b]
    \centering
    \includegraphics[width=.9\linewidth]{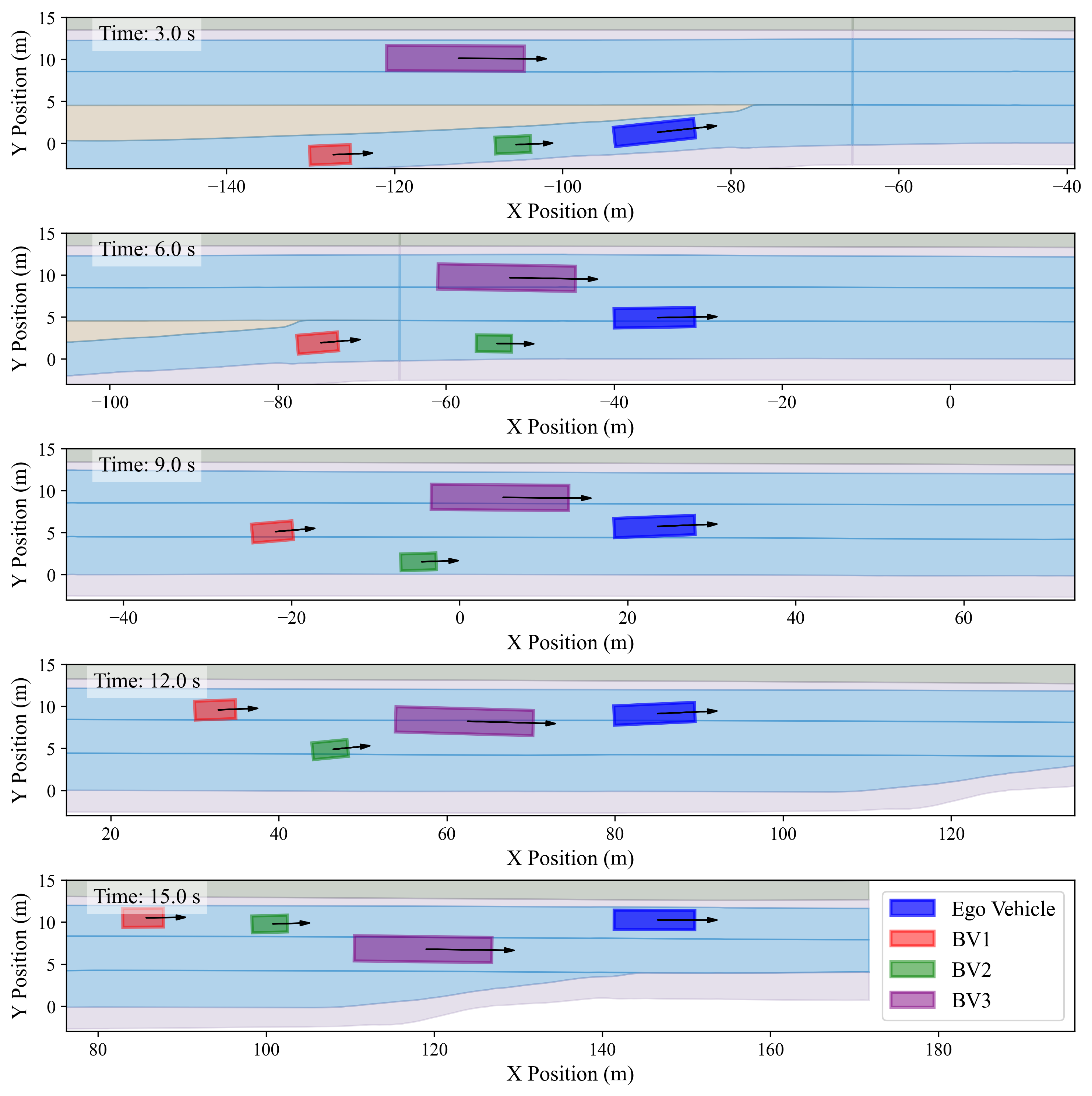}
    \caption{Visualization of the scenario with a large standard deviation in subjective evaluations. (SD=1.34, minTTC=3.74s, maxDNDA=0.49)}
    \label{divergent_sce}
\end{figure}

\begin{figure}[t]%
    \centering
    \subfloat[Participant evaluating the scenario as low risk]{
        \label{low_risk}
        \includegraphics[width=.75\linewidth]{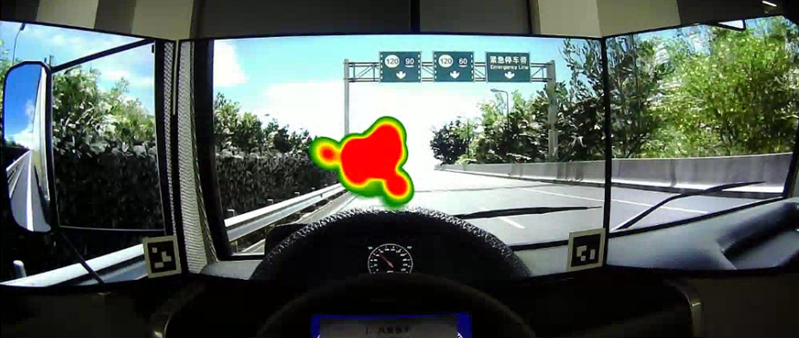}
        } \vspace{0.01cm}
    \subfloat[Participant evaluating the scenario as high risk]{
        \label{high_risk}
        \includegraphics[width=.75\linewidth]{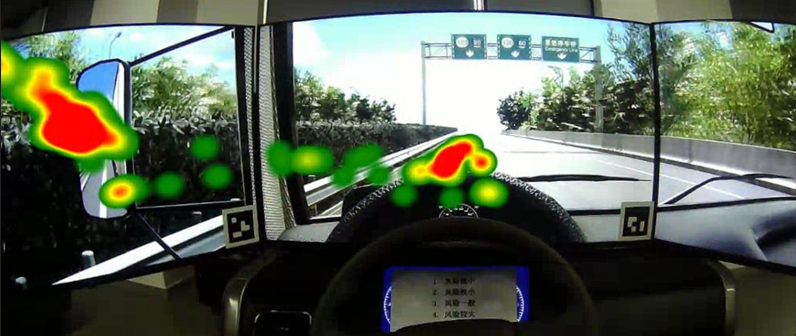}
        }
    \caption{Eye-tracking heatmaps of participants with divergent subjective risk perception in scenario shown in Fig. \ref{divergent_sce}.
    }
    \label{heatmaps}
\end{figure}

For scenarios with large standard deviation in subjective evaluation results, we select an example as visualized in Fig. \ref{divergent_sce}. In this scenario, the ego vehicle first merges from the ramp onto the main road, followed by executing consecutive lane changes to reach the leftmost lane (Lane 1). Throughout these maneuvers, the ego vehicle maintains safety distances from surrounding vehicles with no preceding vehicles in its path, leading a portion of participants to evaluate it as a low-risk scenario. However, given that consecutive lane changes violate Chinese traffic regulations and the ego vehicle exhibits excessive proximity to the left-side guardrail after the lane change, this leads another portion of participants to evaluate it as a high-risk scenario. The above analysis is also well reflected in the participants' eye-tracking heatmaps, as shown in Fig. \ref{heatmaps}. For participant who evaluates the scenario as low-risk, his/her visual attention remains fixated on the front area of the vehicle during the scenario, demonstrating negligible monitoring of rear and lateral areas. In contrast, for participant who evaluates the scenario as high-risk, he/she exhibits comprehensive visual scanning behaviors, actively monitoring not only the front area but also the rearview mirror and lateral forward area. 

In summary, by studying these scenarios with high evaluation consensus or divergence, while combining the participants' eye-tracking data, we can learn different human risk perception patterns, which can provide insights into the development and evaluation of autonomous vehicles.

\section{POTENTIAL APPLICATIONS} \label{sec4}
Since the RISEE dataset contains highly interactive naturalistic driving trajectories and extensive eye-tracking data from diverse human samples, it has great potential for application in both the R\&D and V\&V phases of autonomous vehicles. Here we present several illustrative examples.

\subsection{Risk-aware Personalized Decision-making and Planning}

The RISEE dataset’s subjective risk perception data can be integrated into D\&P systems by serving as risk labels for driving trajectories \cite{yang2023predictiona} or as reward functions via risk assessment indicators \cite{zeng2022riskaware}, enabling human-aligned risk-aware decision-making and planning. Additionally, RISEE’s diverse driver samples (demographics, driving attributes, risk sensitivity) support understanding individualized risk perception patterns, facilitating personalized decision-making and driving style adaptation \cite{jiang2024personalized}.

\subsection{Risk Indicator Construction and Evaluation}

Existing risk assessment indicators primarily rely on current vehicle states and motion predictions, lacking human subjective risk perception incorporating additional deterministic or potential risk factors like lane markings and motion uncertainties in surrounding vehicles \cite{song2024subjective}, which can be identified using the eye-tracking data in RISEE (e.g., repeated gaze on left-front vehicles indicating perceived lane-change risks) for risk indicator constructions. For evaluation, while current studies use collision inevitable time thresholds to assess warning timeliness \cite{yan2024evaluation}, high-level autonomous systems require earlier detection of emerging risks to enable defensive driving. This dataset can benchmark risk indicators by comparing their alerts against human-perceived risk timing in scenarios, evaluating both accuracy and proactive risk anticipation capabilities.

\subsection{Multi-dimensional Driving Intelligence Evaluation}
Existing frameworks for driving intelligence evaluation typically incorporate multiple high-level hierarchies and integrate multi-dimensional fundamental metrics such as safety, comfort, and efficiency \cite{you2025comprehensive}\cite{ma2024evolving}. The driving trajectories with human subjective risk perception provided by the RISEE dataset can effectively assist in training safety performance evaluation models. Meanwhile, thanks to the eye-tracking data in the dataset, participants' pupil diameter, gaze points, and gaze duration can so provide insights into cognitive comfort evaluations \cite{chen2024study}.

\section{CONCLUSIONS} \label{sec5}

In this paper, we present the RISEE dataset, which incorporates highly interactive naturalistic driving trajectories, human subjective evaluations, and their eye-tracking data. To capture realistic and highly interactive traffic scenarios, drone-based traffic videos are first recorded on a highway on-ramp merging section. Subsequently, high-interaction scenarios are manually selected, and simulation reconstructions are performed within simulation software to generate drivers’ FPV videos. To enhance the immersive experience for human evaluators, both the vehicle interior and external environments are visually optimized, and acoustic feedback is also generated. Volunteers are then recruited to watch the FPV videos in a driving simulator, during which their subjective risk perception and eye-tracking data are collected. Analysis of the dataset demonstrates that RISEE contains scenarios with diverse interaction patterns and rich collection of human samples, demonstrating significant potential in both R\&D and V\&V stages of autonomous driving D\&P systems.

\bibliographystyle{IEEEtran}
\bibliography{IEEEabrv,citelist}

\end{document}